# Disk Cooling and Wind Lines as Seen in the Spectral Line Evolution of V960 Mon

Adolfo Carvalho[1], Lynne Hillenbrand[1], and Jerome Seebeck[1,2]
[1] Department of Astronomy, California Institute of Technology, Pasadena, CA 91125, USA
[2] Department of Astronomy, University of Maryland, College Park, MD 20742, USA

## Abstract

We follow up our photometric study of the postoutburst evolution of the FU Ori object V960 Mon with a complementary spectroscopic study at high dispersion that uses time series spectra from the Keck/HIgh Resolution Echelle Spectrograph. Consistent with the photometric results reported in Carvalho et al., we find that the spectral evolution of V960 Mon corresponds to a decrease in the temperature of the inner disk, driven by a combination of a decreasing accretion rate and an increasing inner disk radius. We also find that although the majority of the absorption lines are well matched by our accretion disk model spectrum, there are several strong absorption line families and a few emission lines that are not captured by the model. By subtracting the accretion disk model from the data at each epoch, we isolate the wind and outflow components of the system. The residuals show both broad and highly blueshifted profiles, as well as narrow and only slightly blueshifted profiles, with some lines displaying both types of features.

*Unified Astronomy Thesaurus concepts:* FU Orionis stars (553); Stellar accretion disks (1579); High resolution spectroscopy (2096); Young stellar objects (1834); Stellar winds (1636); Optical bursts (1164)

*Supporting material:* figure set

## 1. Introduction

FU Ori outbursts produce photometric brightenings that reach optical amplitudes of $\Delta V \sim 4$–6 mag, and are associated with episodes of significantly enhanced accretion (Hartmann & Kenyon 1996) in young stellar objects (YSOs). The accretion rates during these outbursts may increase by factors $10^2$–$10^4$, leading to proposals that YSOs may accrete a significant fraction of their mass during the events (see Fischer et al. 2023 for a review).

In 2014 December, the relatively unknown YSO V960 Mon underwent a large outburst, which was initially reported as a suspected FU Ori object by Maehara et al. (2014). V960 Mon has several other YSOs nearby, is surrounded by diffuse dust emission, and its status as an FU Ori object was quickly confirmed by follow-up observations (Hackstein et al. 2014; Hillenbrand 2014; Reipurth & Connelley 2015). The outburst peaked at $V \sim 11.2$ mag and had a relatively flat outburst amplitude across the spectrum, with $\Delta B \sim 3$ reported by Kóspál (2015) and a $\Delta W2 \sim 2.2$ from the Widefield Infrared Survey Explorer (Mainzer et al. 2011). Carvalho et al. (2023; hereafter Paper I) Figure 1 provides the full multiband lightcurve to date.

In the months and years following the outburst, the target faded approximately exponentially, eventually reaching a plateau of $B \sim 14$ ($V \sim 13$) after 2018, as can be seen in Figure 1. Though the target faded rapidly postoutburst, since reaching the plateau in 2018 its brightness has remained essentially unchanged and is still 1.2 mag brighter in the $B$ band than preoutburst.

In Paper I, we presented a disk model that successfully reproduces the color–magnitude evolution of the target during its exponential fade. We used photometry gathered near the outburst epoch and a single high-dispersion spectrum from the same time to determine the best-fit system parameters in the pure-accretion scenario. The stellar and disk parameters that best explain the outburst peak are $M_* = 0.59\ M_\odot$, $R_{\mathrm{inner}} = 2.11\ R_\odot$, and $\dot{M} = 10^{-4.59}\ M_\odot\ \mathrm{yr}^{-1}$, corresponding to $T_{\max} = 8240$ K, $L_{\mathrm{acc}} = 113\ L_\odot$, and $v_{\max} = 60$ km s$^{-1}$. Our analysis in Paper I assumed a distance to the target of 1120 pc (Kuhn & Hillenbrand 2019).

The model described in detail in Paper I is used here to study the spectral evolution of the V960 Mon system at high dispersion. We followed the exponential decline over several years, from the initial outburst to the beginning of the plateau, as indicated in Figure 1. In this paper, we present the spectra, gathered with the Keck/HIgh Resolution Echelle Spectrograph (HIRES), and demonstrate that our accretion disk model is able to reproduce the spectral evolution accurately over a broad range of optical wavelengths during the fade.

We then isolate the excess absorption and emission in the spectrum by subtracting our high-resolution model from the data. In this way, we are able to analyze a spectrum of the nondisk components in the system.

We begin by discussing our reduction and continuum normalization of the HIRES spectra in Section 2. We then give a brief summary of our disk model from Paper I, followed by a discussion of the spectral line evolution in the system in Section 3. We show evidence for a cooling inner disk and how that is traced by the absorption lines in the HIRES spectra (and reproduced by our model) in Section 4. Once we have subtracted the model spectra from the data, we analyze the excess absorption and emission spectrum, which is presented in Section 5. We identify several forbidden emission lines in the spectrum, which grow as the target fades, and we show those in Section 6. We discuss our results in the context of existing FU Ori object and other young star literature in Section 7 and summarize our conclusions in Section 8.

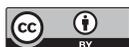

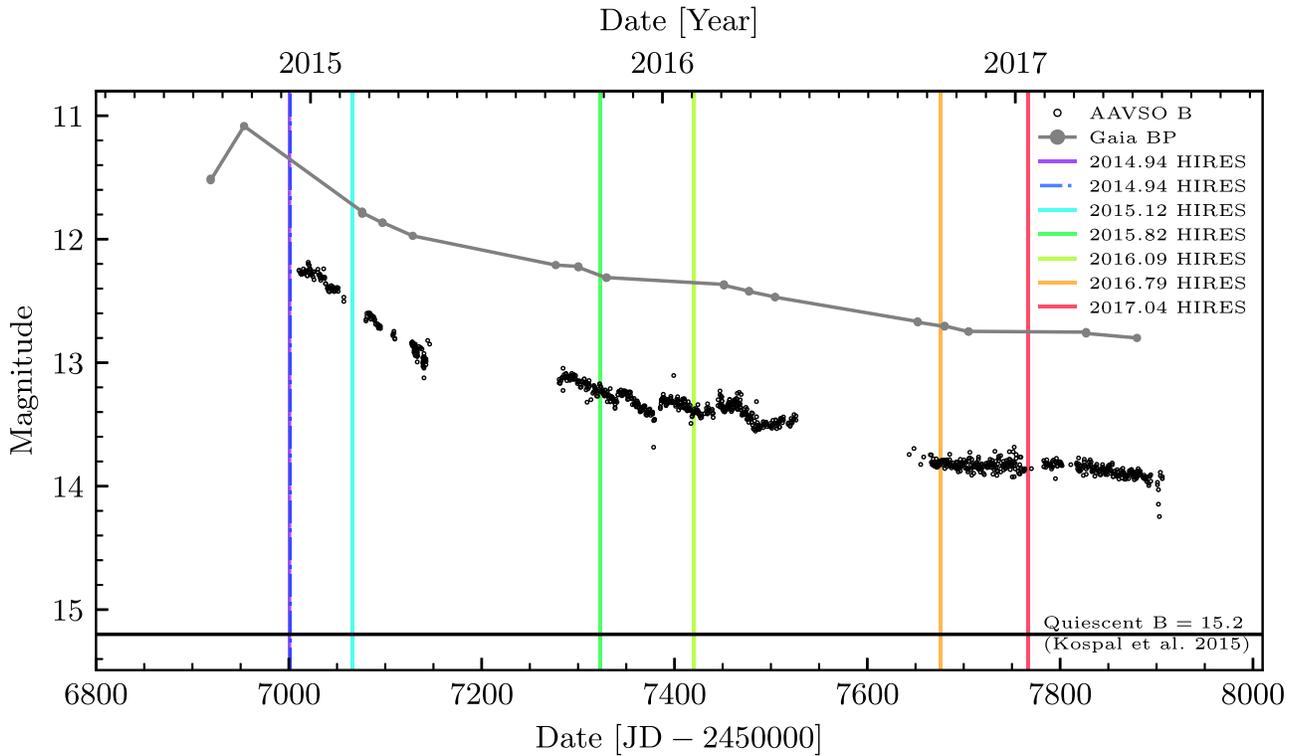

**Figure 1.** The epochs of our HIRES spectra are shown (vertical solid lines) relative to the AAVSO B (black circles) and Gaia BP (gray connected points) lightcurves of V960 Mon. The lightcurve illustrates the rapid postoutburst fading and eventual plateauing in later epochs. The quiescent B magnitude reported in Kóspál (2015) is shown as the black horizontal line for reference. The 2014 December 10 HIRES epoch is marked with a dotted–dashed line due to overlap with the 2014 December 9 epoch.

## 2. Data

We obtained visible range high-dispersion spectra from the Keck Observatory's HIRES (Vogt et al. 1994), covering 4780–9220 Å. Table 1 gives the epochs, instruments, and signal-to-noise ratios (S/Ns) for the spectra. The spectra were processed with the 2008 version of the MAKEE pipeline reduction package written by Tom Barlow.[3]

We normalize the spectra by fitting the continuum using a regularized asymmetric least-squares technique (Eilers & Boelens 2005). The regularization parameter allows for more or less flexible continuum fits and the technique is more robust to the edges of the spectrum than polynomial fitting. Orders with emission lines (e.g., Hα, weak forbidden emission lines, and the Ca II infrared triplet (IRT)) need special treatment. We mask emission lines in the spectrum and use the linear interpolation from the redward continuum point to the blueward continuum point on either side of the lines as the continuum under those lines. Several orders from the continuum-normalized HIRES outburst epoch spectrum are illustrated in Figure 2.

As is mentioned in Paper I, we compute the half-width at half-depth (HWHD) of several absorption lines in the outburst spectrum (taken 2014 December 9) across the optical range, and find no correlation between wavelength of the line and HWHD. The measurements are discussed and shown in Section 3.2. The mean and standard deviation of the measurements are $44 \pm 5$ km s$^{-1}$, consistent with the line width measurements reported by Park et al. (2020).

We compute the equivalent widths (EWs) of select lines via direct integration, taking the continuum to be 1.0 following our normalization described above. The procedure for the EW measurement and uncertainty estimation is described in detail in Carvalho & Hillenbrand (2022) and our results for V960 Mon are presented in Section 4.

---
[3] https://sites.astro.caltech.edu/~tb/makee/

**Table 1**
Spectroscopic Observations Log

| Date | Instrument | S/N at 7100 Å | Exposure Time (s) |
| --- | --- | --- | --- |
| 2014-12-09 | HIRES | 170 | 600 |
| 2014-12-10 | HIRES | 117 | 315 |
| 2015-02-09 | HIRES | 98 | 245 |
| 2015-10-27 | HIRES | 97 | 300 |
| 2016-02-02 | HIRES | 147 | 600 |
| 2016-10-14 | HIRES | 80 | 180 |
| 2017-01-13 | HIRES | 52 | 180 |

We measure a heliocentric systemic velocity of $+43.0$ km s$^{-1}$, which is roughly consistent with the $v_{LSR} = 23.81$ km s$^{-1}$ ($v_{helio} \sim 40$ km s$^{-1}$) measured by Cruz-Sáenz de Miera et al. (2023).

The seven epochs of HIRES data, along with the residuals from the high-dispersion disk model described below, are shown in Figure 2.

## 3. Modeling the High-dispersion Spectra

We use the disk model described in Paper I to model the high-dispersion data. We briefly summarize the model below, as well as the technique we adopt to model the evolution of the system from outburst to later epochs. We describe how temperature-sensitive lines behave in the model in terms of their presence and broadening We also discuss the effect of varying the accretion rate and innermost radius on the high-dispersion models and the strong agreement with what we observe in the HIRES spectra.

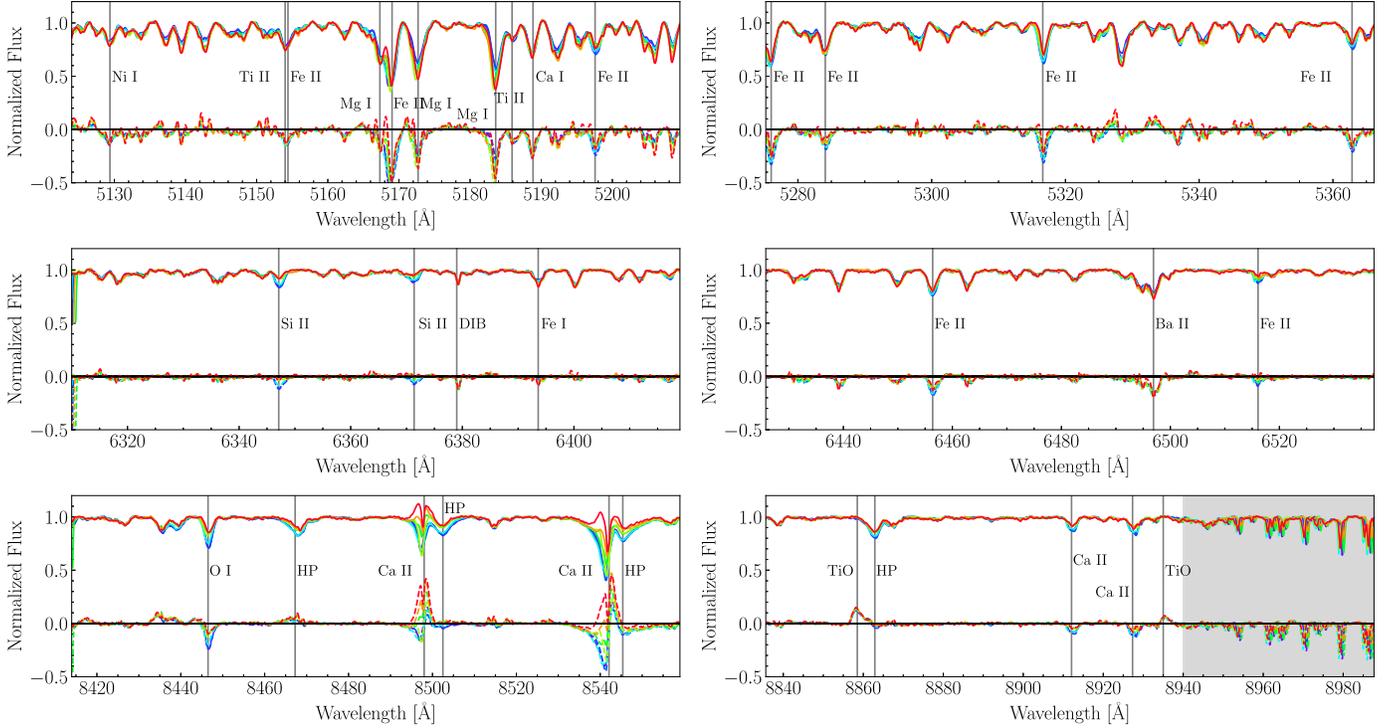

**Figure 2.** The HIRES spectra shown in a time series, with bluer to redder spectra indicating earlier to later epochs. The lower set of curves in each panel are the residuals after subtracting the disk model appropriate to each epoch. See Figures 2–4 of Paper I for a direct comparison between the outburst epoch spectrum (bluest line here) and the disk model. Strong excess absorption in certain lines (marked) is attributed to wind contributions in the spectra; see Section 5. Outside of these lines, the typical rms value of the residuals is <3%. The remaining 28 orders are shown in Appendix C.

### 3.1. A Recap of the Disk Model

We assume that the system can be well approximated by a thin accretion disk, following the Shakura & Sunyaev (1973) model, with the change that for radii interior to that of the maximum temperature, $T_{max}$, we impose a flat profile, following Kenyon et al. (1988). Therefore, the radial temperature profile in the model is given by:

$$T_{eff}^4(r) = \frac{3GM_*\dot{M}}{8\pi\sigma r^3}\left(1 - \sqrt{\frac{R_{inner}}{r}}\right), \quad (1)$$

for $r \geqslant \frac{49}{36} R_{inner}$ and $T_{eff}^4(r) = T_{max}$ for $r < \frac{49}{36} R_{inner}$. Here, $M_*$ is the mass of the central star, $\dot{M}$ is the accretion rate, $R_{inner}$ is the innermost radius of the accretion disk, $\sigma$ is the Stefan–Boltzmann constant, and $G$ is the gravitational constant.

We use PHOENIX (Husser et al. 2013) model atmosphere spectra[4] corresponding to the $T_{eff}$ of each annulus of the disk, following the temperature profile in Equation (1). We find that using only the log $g = 1.5$ atmospheres gives a better match to the high-dispersion spectra than the $g(r)$ model implemented in Paper I. The change does not affect the spectral energy distribution (SED) fits from Paper I.

To account for turbulence in the disk like that seen in the simulation of FU Ori presented by Zhu et al. (2020), we apply 20 km s$^{-1}$ of spherical broadening to the atmospheres using the direct integration method in Carvalho & Johns-Krull (2023). The broadening is similar to stellar rotational broadening, but without limb darkening. This initial broadening step helps to match the individual line profiles better, which do not show the narrow double peaks typical of Keplerian rotation in a thin disk, but instead have a flat-bottomed, box-like profile. We then apply the disk Keplerian broadening as described in Paper I.

---
[4] Downloadable at http://svo2.cab.inta-csic.es/theory/newov2/index.php.

In Paper I, we found that as the target fades, the color–magnitude evolution is well matched by varying both $\dot{M}$ and $R_{inner}$. The exact scaling between the two quantities is that given by the canonical truncation radius equation (Equation (6) of Paper I), which yields $R_{inner} \propto \dot{M}^{-2/7}$. We use the color–temperature calculated in Paper I at each HIRES epoch from the AAVSO photometry to estimate the appropriate $\dot{M}$, assuming the scaling $T \propto \dot{M}^{13/28}$. We compute a $T_{max}$ at each epoch ranging from 8300 to 6600 K. See Figure 11 of Paper I for the resulting temperature profiles.

The high-dispersion disk models computed from the light-curve evolution are generally very good fits to the HIRES spectra. The models reproduce the absorption lines across the entire spectral range well, with the exception of certain features we believe trace nondisk absorption and emission components (including "wind" lines like H$\alpha$ and Ca II, and high-excitation potential (EP) lines like Si II and C I, see Section 5). The typical rms of the residuals, excluding these and other key excess features that are not accounted for in the disk model, is <3%.

### 3.2. Temperature-sensitive Lines: Differential Broadening

In the canonical FU Ori disk model (Kenyon et al. 1988; Calvet et al. 1993), the spectral line broadening is given by the Keplerian rotation of the gas disk. Therefore, for a given spectral line, the broadening should be proportional to $\sqrt{GM_*/r_{\rm line}}$, where $r_{\rm line}$ is the radius in the disk where the line is expected to form. For lines with higher EP, one might expect $r_{\rm line} \sim R_{\rm inner}$, whereas lower-EP lines are expected to form further out in the disk, on average. While this may be the case, we also find that the $\dot{M}$ in the model plays a role in determining the final observed line broadening.

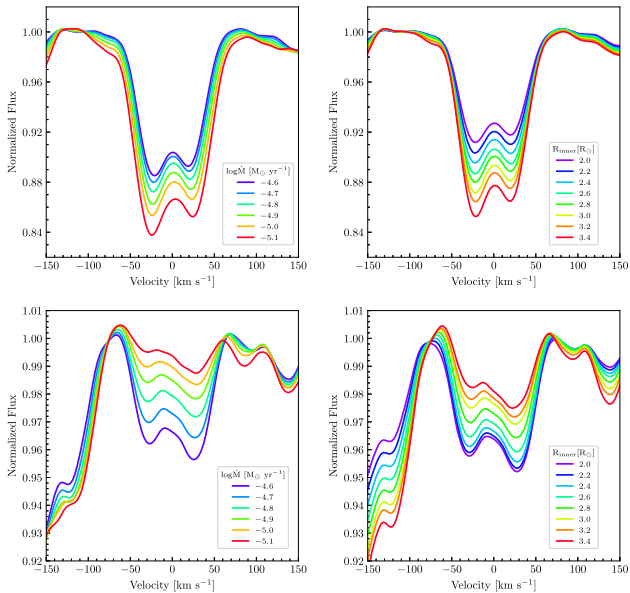

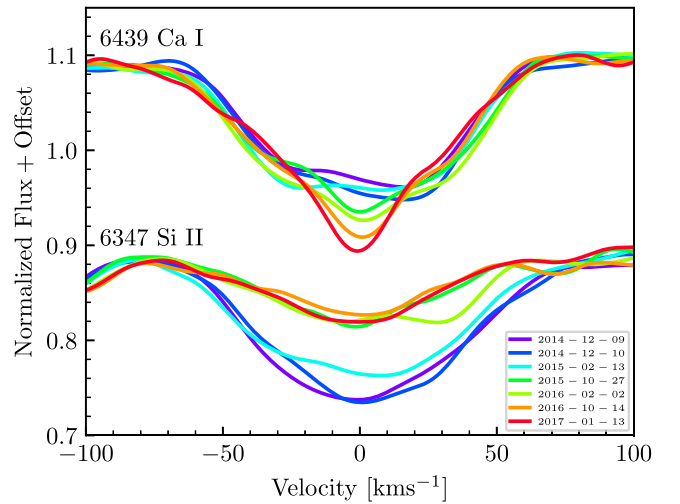

**Figure 3.** The effect of varying $\dot{M}$ and $R_{\rm inner}$ on the width and depth of absorption lines with different EPs. Left: models of the Ca I $\lambda 6439$ (upper panel) and Si II $\lambda 6347$ (lower panel) lines for different values of $\dot{M}$. Notice the Ca I $\lambda 6439$ line broadens as $\dot{M}$ decreases, whereas the Si II $\lambda 6347$ line only changes in depth. Right: models of the same two lines for different values of $R_{\rm inner}$. Notice here that the $T_{\rm max}$-dependent broadening still dominates in the Ca I line as $R_{\rm inner}$ decreases, making the line counterintuitively broader. However, in Si II, the broadening decreases with increasing $R_{\rm inner}$, as expected due to the decrease in the maximum Keplerian velocity.

**Figure 4.** The Ca I $\lambda 6439$ (upper profiles) and Si II $\lambda 6347$ (lower profiles) absorption lines shown for the different epochs of the HIRES spectra. Notice the Si II line becomes narrower, in addition to weaker, consistent with an increase in $R_{\rm inner}$, while the broadening of the Ca I line remains relatively unchanged.

We find that for lower-EP lines belonging to neutral species such as Fe I and Ca I, lower values of $\dot{M}$ produce broader line profiles and higher $\dot{M}$ values produce narrower line profiles. In fact, the effect is so strong in these lines that it overwhelms the effect of varying $R_{\rm inner}$, as shown in Figure 3.

This is because for higher values of $\dot{M}$, the $T_{\rm eff}$ in the fastest-moving annuli is high enough that low-EP lines from Ca I (e.g., Ca I $\lambda 6439$) and Fe I (e.g., Fe I $\lambda 6393$) are extremely weak or totally absent. The effect is especially pronounced in the $T_{\rm eff} > 7000$ K annuli. In this case, the low-EP lines will not be broadened as significantly as lines that still appear in those hottest annuli, such as Si II $\lambda 6347$ and $\lambda 6371$.

As $\dot{M}$ decreases, we see in Figure 3 that the Ca I $\lambda 6439$ line grows broader because the fastest-moving, closest-in annuli are cool enough to show lower-EP lines in absorption. The higher-EP lines like Si II $\lambda 6347$ remain the same width throughout, because the innermost annuli do not get hot enough for the lines to disappear. Decreasing $\dot{M}$ only decreases the depth of the lines. This is clear in the lower left panel of Figure 3.

We see then, that this dependence on $\dot{M}$ in the line broadening is in fact a dependence on $T_{\rm max}$. This implies that changes to $R_{\rm inner}$ should elicit the same effect. Varying $T_{\rm max}$ via $R_{\rm inner}$, as we propose in Paper I, however, is expected to affect the rotational broadening of all lines directly by changing the maximum Keplerian velocity in the disk. The question becomes: are changes in the rotational broadening of lines due to changes in the maximum Keplerian velocity distinguishable from those due to changes in $T_{\rm max}$?

The two panels in the right column of Figure 3 show our investigation of this in the Ca I $\lambda 6439$ and the Si II $\lambda 6347$ lines. In the Ca I line, the $T_{\rm max}$ effect dominates, working to broaden the line as $R_{\rm inner}$ increases due to the overall decrease in $T_{\rm max}$. In the Si II line, the decreasing maximum Keplerian velocity dominates and we see the line narrows as we increase $R_{\rm inner}$.

We find this is consistent with the broadening we see in the high-dispersion spectra. Figure 4 shows the Si II and Ca I lines over time as observed in the HIRES spectra at different epochs. The Ca I line remains at a relatively constant width, tending toward being slightly broader at later epochs. This is what we expect from a decreasing $\dot{M}$ and increasing $R_{\rm inner}$, as seen in Figure 3. The Si II line narrows rapidly toward later epochs, as we might expect from the discussion above and as is shown in Figure 3. As we discuss in Section 5.2, the Si II line does not arise entirely from the disk, so its rapid narrowing is not only due to the increase in $R_{\rm inner}$.

### 3.3. Line Broadening as a Function of Wavelength

HWHD measurements have been used by many authors to argue for the presence (Welty et al. 1990; Park et al. 2020) or absence (Herbig et al. 2003) of disk-broadened spectral lines in FU Ori stars. However, due to the differential broadening of different spectral lines based on their location of formation in the disk, as just discussed (Section 3.2), it is not straightforward to connect an HWHD versus wavelength (or even EP) relation to the physics of the disk.

We measured the HWHD values for several lines in the observed outburst spectrum and in outburst epoch model spectrum of V960 Mon (shown in Figure 5). To estimate the uncertainty of the HWHD measurements, we fit disk profiles to the lines and multiplied the fractional $1\sigma$ uncertainty in the width parameter of the fit by our HWHD value. While there is a generally decreasing (though not statistically significant) trend in HWHD versus wavelength over broad regions of the spectrum, there is significant scatter in the measurements of even neighboring lines. This is in agreement with the results presented in Zhu et al. (2009), which also show that for FU Ori, the disk model predicts almost no wavelength dependence on the HWHD measurements made in the optical.

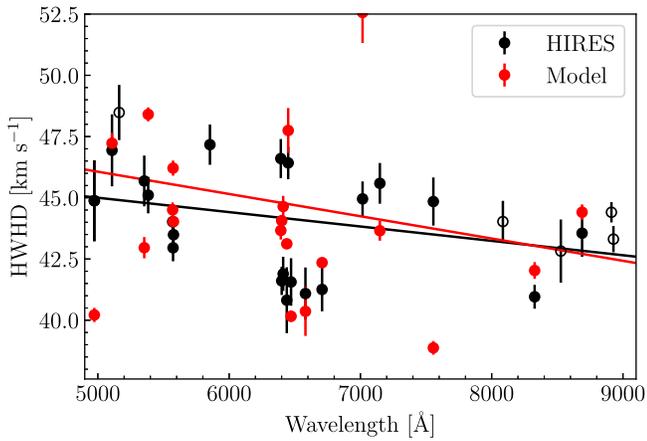

**Figure 5.** The HWHD measurements of several isolated absorption lines in the HIRES outburst (2014 December 9) spectrum (black) and the spectral model (red). Empty symbols show lines that are isolated and easily measured in the data but appear severely blended in the models. Although there are some line-by-line inconsistencies (largely due to imperfections in the PHOENIX models), neither set of measurements is significantly correlated with wavelength. The data and the disk model also have similar standard deviations and the scatter in the HWHD measurements does not vary with wavelength. The error bars shown reflect the uncertainty in the width parameter fit used to compute the HWHD measurements.

Both the slightly decreasing trend with wavelength and the large scatter are consistent with the high-resolution disk model, as shown in Figure 5. The HWHD measurements from our high-resolution model have a slightly lower median than those in the HIRES data. However, attempting a model with slightly greater median broadening gives a worse fit to the data, with greater chi-squared values. For a more direct comparison between the two data sets, we scale the model HWHD values to have the same median as the HIRES values.

Linear fits to the two HWHD versus wavelength relations give slopes of $-3.8 \times 10^{-4}$ km s$^{-1}$ Å$^{-1}$ and $-9.3 \times 10^{-4}$ km s$^{-1}$ Å$^{-1}$ with low-significance Pearson test $p$ values of 0.33 and 0.40 for the data and model, respectively. The scatter of 2.1 km s$^{-1}$ in the HIRES HWHD measurements is greater than any wavelength-dependent change in width predicted by the disk model over the 5000–9000 Å wavelength range.

## 4. Evidence of Disk Cooling in the High-dispersion Spectra

We look to the behavior of the temperature-sensitive absorption lines in the high-dispersion spectra to confirm the temperature evolution of the disk that is seen in the photometric models presented in Paper I. We will focus on two sets of lines: those with very high EP values (>7 eV) and those with relatively lower-EP values (<3 eV). In general, we find that the high-EP lines weaken rapidly as the target fades, while the low-EP lines become deeper over the same time. This is expected for a target that is cooling, as the higher energy levels depopulate and fill the lower energy levels.

One challenge to interpreting the behavior of the highest-EP lines in the HIRES spectra is that some lines are significantly stronger than the models predict them to be. We believe this is evidence that the highest-EP lines, such as Si II λ6347 and λ6371, O I λ8446, Ca II λλ8912, 8927, and C I λ9111, trace a nondisk (potentially outflowing) component in the system. We will discuss these lines in detail in Section 5.2. The high-EP lines we will focus on in this section are those that do not appear to be influenced by wind.

To illustrate the general time evolution of the lines in our HIRES spectra of V960 Mon, we have chosen four lines to highlight: two high-EP and two low-EP lines. The high-EP regime is represented by the Paschen series lines of H I (hereafter HP), with the HP λ8862 line serving as a specific example, and the Fe II λ5316 line. These are both isolated features that show dramatic decreases in line strength as the target fades. The low-EP lines are represented by the Fe I λ5328 line and the Ca I λ6439 line. The Fe I λ5328 feature (which is in reality a blend of low-EP Fe I lines) is especially temperature sensitive, increasing in strength dramatically as the target fades (see Figure 2). The Ca I λ6439 line has the advantage of being isolated, so it clearly shows the flat-bottomed line profile characteristic of FU Ori objects (though the profile develops a rest-velocity excess absorption feature over time, as discussed in Section 5.3). It also grows significantly in time, as seen in the profiles shown in Figure 4.

We quantify the weakening or strengthening of lines we will discuss in this section by computing their EWs at each HIRES epoch. The measurements are shown in Figure 6 with the V-band lightcurve from Paper I plotted alongside the measurements as a reference for the brightness evolution of the target. In the figure, we see that the shape of the EW curves qualitatively either follows or mirrors that of the V-band lightcurve, depending on the line. The high-EP HP line and the Fe II λ5316 line both show decreasing EW measurements over time. The HP line closely follows the lightcurve, with a greater slope matching the initial rapid fade of the target, then plateauing at later epochs. The lower-EP lines, Fe I λ5328 and Ca I λ6439, both mirror the lightcurve, growing rapidly during the early epochs and plateauing at larger values later.

### 4.1. The Time Evolution of the HP Series

As can be seen in Figure 7, our disk model is a good match to the line strength of the HP λ8862 line at outburst and its subsequent time evolution. We interpret this to mean three things: the HP lines are good tracers of disk temperature; our outburst model $T_{max} \sim 8300$ K is sufficiently high to describe the initial HP depth; and the line evolution supports our proposed temperature evolution of the disk. We note that the HP lines blueward of 8350 Å are not detected, which is consistent with our model predictions, due to blending with other features and minimal time evolution. Of the lines blueward of 8350 Å, only the 8345.5 Å line varies by more the ~1%.

The exponential decrease of the HP λ8862 line is a strong indication that the line is closely tracing the decrease of $T_{max}$, which is driving the brightness decrease in V960 Mon. We note that this differs somewhat from the behavior of the Fe II λ5316 line, which also decreases in strength, but shows a more linear decline in time. The difference in evolution can be explained by the relative temperature sensitivities of the two lines. The HP lines are strongest in very hot atmospheres and will weaken quickly as the hottest components of the disk cool. This makes the HP lines very sensitive to the $T_{max}$ in the disk. The Fe II lines span a broader range of temperatures and are relatively temperature insensitive in our disk models. Therefore, as the disk cools, the EW of the Fe II lines may decrease but the annuli in which the line largely forms will simply shift radially inward, slowing the weakening of the line.

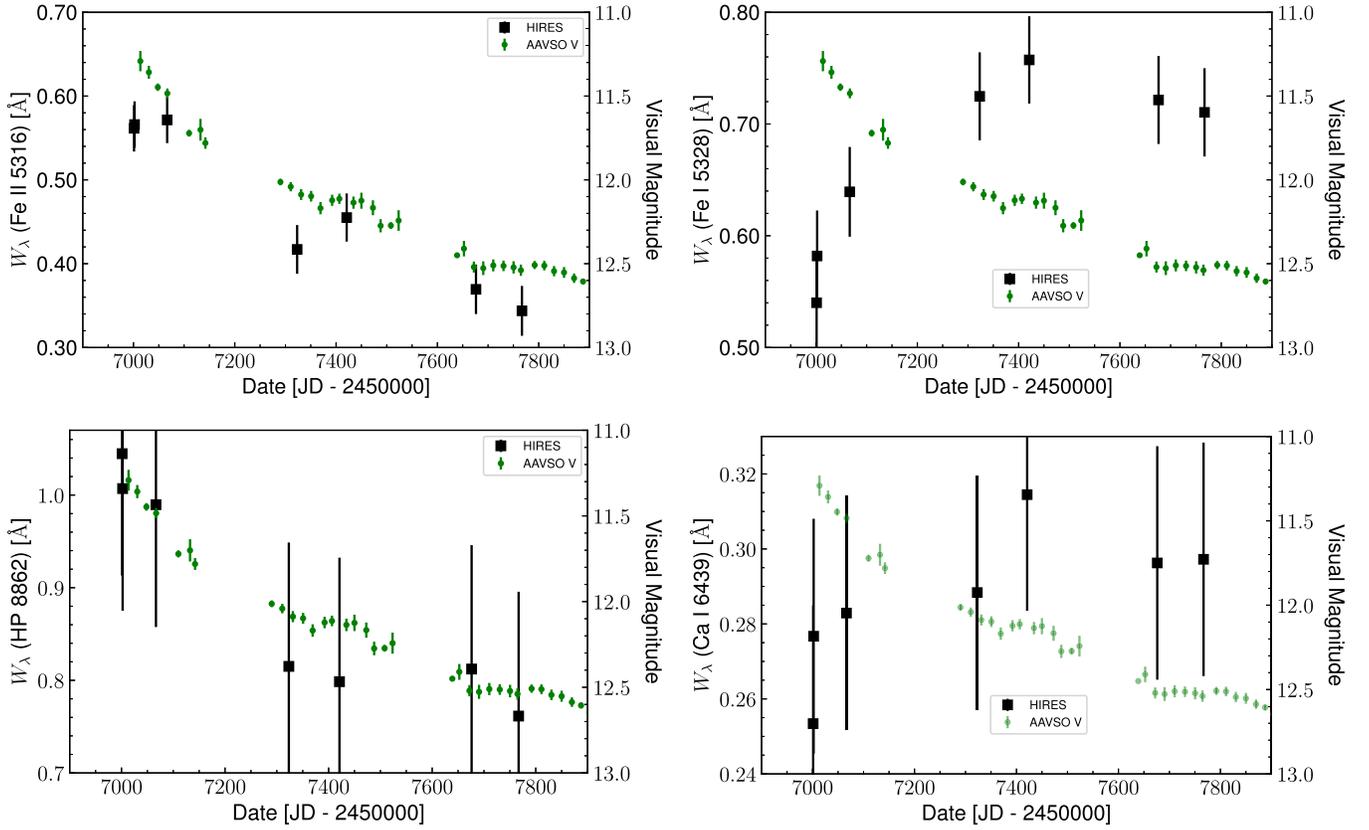

**Figure 6.** The EW evolution (black squares) of two higher-EP and two lower-EP lines in the spectrum of V960 Mon. The AAVSO *V*-band lightcurve is plotted (green points) for reference, showing the evolution of the continuum brightness over the spectral epochs. Left column: the evolution of the $\lambda$5316 Fe II line and $\lambda$8862 HP EWs, which both closely follow the lightcurve of the target in their decreasing strength. The EW measurement for the $\lambda$8862 HP line is blended with the Fe I $\lambda$8866 line, but the Fe I line does not vary in time. Right column: the evolution of the $\lambda$5328 Fe I line and $\lambda$6439 Ca I line, where both follow an inverse trend to the lightcurve in their growth, similarly strengthening as the target fades and plateaus at later times.

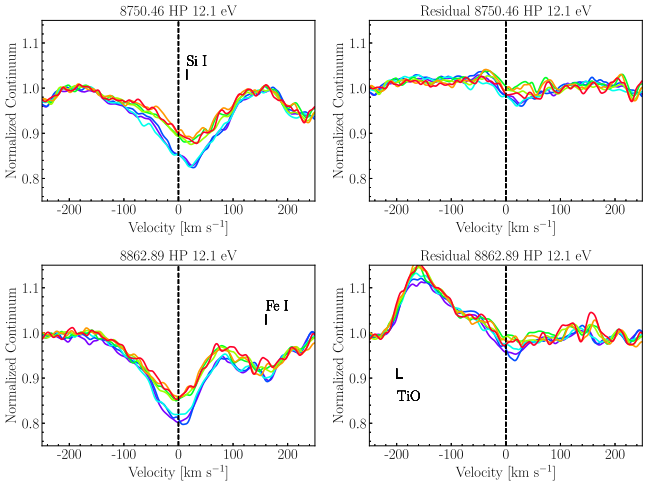

**Figure 7.** The 8750 and 8862 Å HP lines in the HIRES spectra (left column) and the residual spectrum (right column). Notice the lines have been removed to within a few percent in every epoch, indicating the modeled $T_{max}$ change reproduces the evolution of the HP series accurately. The 8750 Å line is blended with the $\lambda$8751 Si I line and the 8862 Å line is blended with an Fe I line at 8866 Å but both are well modeled, do not show any evolution, and do not appear in the residuals. The emission residual in the lower right panel at $-200$ km s$^{-1}$ is due to TiO absorption in the model that is not present in the data. We discuss this discrepancy in Appendix B.

### 4.2. The Time Evolution of the Low-EP Metal Lines

We also see evidence of disk cooling in the behavior of temperature-sensitive low-EP lines, such as Fe I $\lambda$5328 and Ca I $\lambda$6439. The region of the disk from which the optical continuum arises is at 5000–7000 K, which is hot enough that the majority of the Ca and Fe is ionized. This makes the Ca I and Fe I lines particularly sensitive temperature tracers (Gray 2008).

One supporting argument for the cooling disk was presented in Section 3.2, where we attribute the slight broadening of the Ca I $\lambda$6439 line in time to cooling of the innermost annuli. Another argument is that the Fe I and Ca I lines increase in strength significantly as the target fades. The line strength increase is seen clearly in the EW measurements of the Fe I $\lambda$5328 and Ca I $\lambda$6439 lines in Figure 6.

In order to constrain the disk properties from the EW measurements of the low-EP lines more directly, we compute the EW ratios of these lines to neighboring lines. Using EW ratios allows us to investigate the local gravity and temperature from the regions where the lines were emitted without worrying about contributions from continuum opacity.

We focus our analysis on two line ratios in particular: the $\lambda$6439/$\lambda$6456 ratio, between the Ca I $\lambda$6439 line and the Fe II $\lambda$6456 line, and the $\lambda$5328/$\lambda$5316 ratio, between the Fe I $\lambda$5328 line and the Fe II $\lambda$5316 line. The denominators of the line ratios are chosen for two reasons: in the case of Fe II $\lambda$6456, the line remains unchanged in the spectra, allowing a good comparison with the evolution of the Ca I $\lambda$6439 line, whereas for Fe II λ5316, its evolution is opposite that of Fe I λ5328, potentially making the ratio more temperature sensitive.

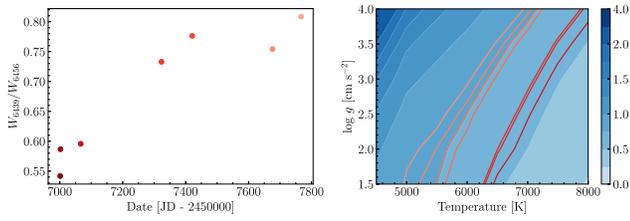

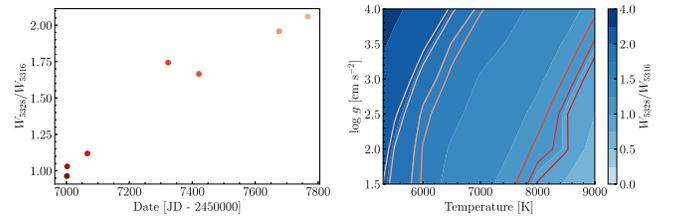

**Figure 8.** Left: the EW ratios of the Ca I λ6439 line and the Fe I λ6456 line as a function of time in the HIRES spectra. Right: the EW ratios of the same lines in the PHOENIX grid in the blue background and the ratios in the HIRES spectra shown in the red color map, with larger EW ratio values shown as darker shades.

**Figure 9.** Left: the EW ratios of the Ca I λ6439 line and the Fe I λ6456 line as a function of time in the HIRES spectra. Right: the EW ratios of the same lines in the PHOENIX grid in the blue background and the ratios in the HIRES spectra shown in the red color map, with larger EW ratio values shown as darker shades.

We study these two line ratios by comparing to computed expected line ratios in the PHOENIX grid, for a range of temperatures (5000–9000 K) and range of gravities (1.5 ⩽ log $g$ ⩽ 4.0). This gives us an EW ratio surface, on which we can plot the EW ratios we compute in the HIRES spectra. Placing the EW measurements on their appropriate contours shows the corresponding temperature and gravity for that ratio in the PHOENIX grid.

The resulting contour plots are shown in Figures 8 and 9, along with the time series of the ratios for reference. As predicted from our model, both sets of ratios are initially consistent with relatively high temperatures and evolve toward cooler ones. In fact the λ5328/λ5316 ratio is a good match to the predicted $T_{max}$ evolution of the system.

Ultimately, both the high-EP HP lines and the low-EP neutral atomic lines show good agreement with our model of a cooling disk. Both sets of lines also support our $T_{max}$ estimate for the outburst epoch, particularly in the fact that the HP line depths are well matched for that epoch and those that follow.

## 5. Lines with Excess Absorption Relative to the Disk Model

In the high-dispersion spectra, there are three distinct families of features which are not well matched by our pure-accretion disk models. This section highlights lines showing high-velocity, blueshifted excess absorption; then lines showing broad, rest-velocity absorption that shrinks over time; and lines showing a narrow, rest-velocity absorption that grows over time.

The high-velocity blueshifted features include lines which typically trace winds in young stars and show absorption velocities as high as −200 km s$^{-1}$. These may probe the wind/outflow acceleration region in the disk. We find that the broad, rest-velocity absorption in high-EP lines may trace a hot wind, with $T \sim$ 8000–9000 K. The narrow, rest-velocity absorption in lower-EP lines may in turn trace the outflow as it cools. We describe the line families individually below and discuss possible physical mechanisms from which they may arise.

### 5.1. Blue Excess Absorption

There is evidence of a fast-moving and evolving outflow in the traditional wind-tracing lines in the V960 Mon spectra. We identify several morphological features in the Hα, Hβ, Na I D, and Ca II IRT lines, shown in Figures 10 and 11. The lines trace different velocity components in the flow and the variations in those components over time.

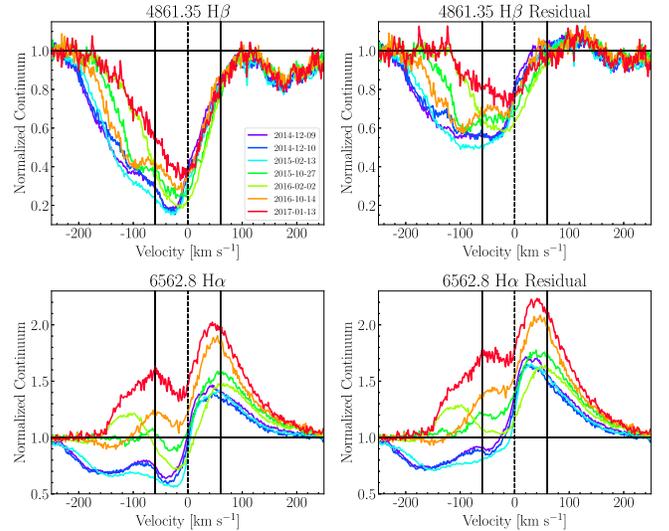

**Figure 10.** The Hβ and Hα lines in each of the HIRES epochs as seen in the original data (left) and as residuals after removing the disk contribution (right). Both lines initially show a high-velocity blueshifted absorption component that weakens rapidly over time and becomes an emission feature in Hα. The profiles also contain a slow component that moves redward. The emission components of the Hα profile are both peaked at ±60 km s$^{-1}$ and strengthen consistently relative to the continuum. The absorption in the upper left panel at +185 km s$^{-1}$ is the Cr II 4864 Å line. Notice it is well removed by the model and is much weaker in the upper right residual panel.

In the outburst epoch and the two epochs shortly after outburst (∼1 month after) we see a strong, blueshifted absorption component in Hα and Hβ extending to −200 km s$^{-1}$. The component is strongest in Hα and disappears sometime between 2015 February and 2015 October, over which time the source faded by 0.8 mag in the B band. There is a distinct component seen in Hβ with an absorption peak at −110 km s$^{-1}$ that disappears between 2015 October and 2016 February, but then reappears during the 2016 October epoch. The feature again disappears by the 2017 January epoch and is not apparent in the later spectra shown in Park et al. (2020).

There is also a slower absorption component in the H lines, which shows an absorption peak around −30 km s$^{-1}$ in both Hα and Hβ at the outburst epoch, but moves to slower velocities at later times. In the 2017 January epoch, the component has slowed to −10 km s$^{-1}$.

The Hβ line has a redshifted absorption wing that is similar in slope to the disk-tracing atomic features. That the absorption likely arises in the disk is confirmed by the fact it is almost fully removed in by our disk-model subtraction, as seen in the right column of Figure 10. The blueshifted absorption features

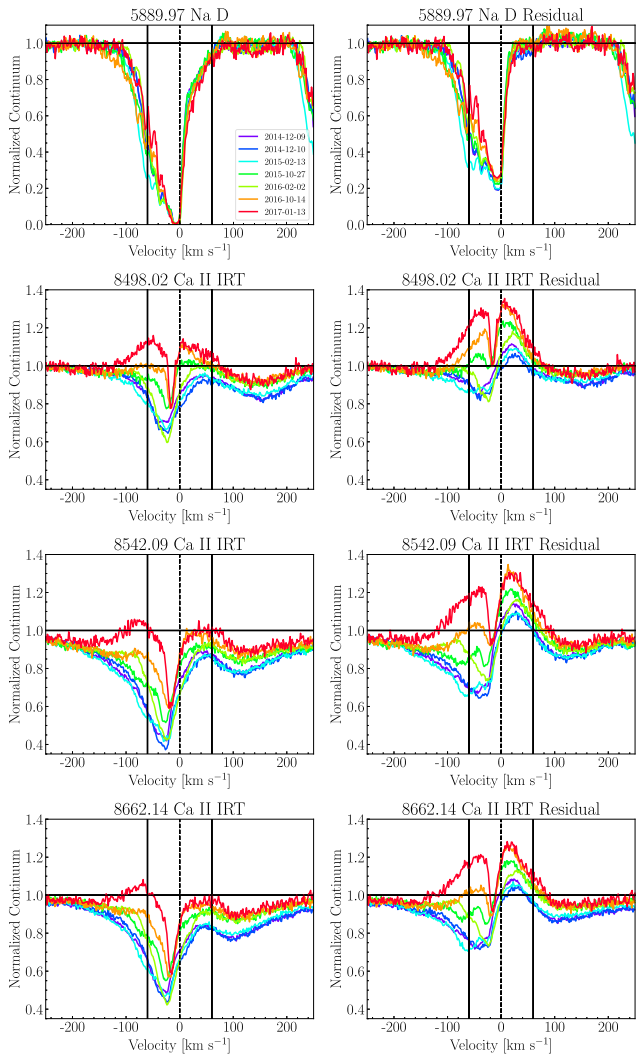

**Figure 11.** The Na D $\lambda5889$ line (top row) and Ca II IRT lines (lower three rows), shown in the data (left column) and residuals (right column). The Ca II IRT lines show a weakening fast blueshifted component and a $\pm60$ km s$^{-1}$ emission component similar to those seen in the H lines. Note that the $+100$ km s$^{-1}$ absorption is due to HP line blending with the IRT. This is well matched by our disk model and mostly removed in the residuals column. The Na D lines are saturated from $-10$ to 0 km s$^{-1}$, which matches the narrow absorption seen in the latest epoch, indicating there may be a constant slow wind covering almost the entire visible emission region of the disk.

discussed above persist in the residuals, indicating they are indeed tracing outflow components.

The especially fast absorption in the H lines is absent in the other wind tracers. The slower component, however, is visible in both H lines and the Ca II IRT. At the outburst epoch, there is a distinct absorption minimum at $-30$ km s$^{-1}$, which slows over time to $-10$ km s$^{-1}$. The change in velocity of this absorption line is most noticeable in the Ca II IRT profiles. We note also that the Ca II IRT profiles appear quite different from one another in the HIRES spectra, despite being a triplet (Figure 11, left column). We find that this is due to the differing levels of HP blending each line experiences, and when the disk (and therefore HP) contribution is subtracted, the lines show similar profiles, as expected (Figure 11, right column).

The Na D lines are remarkably featureless compared with the other wind lines. They are saturated from $-10$ to 0 km s$^{-1}$, indicating the presence of a slow wind that covers the entire optical continuum emission region. The blue wings of the lines extend to $-100$ km s$^{-1}$, tracing a much slower component than the H$\alpha$ and H$\beta$ lines. There is also a component at $-60$ km s$^{-1}$ that is slightly deeper at earlier epochs, but the change is relatively small compared to that in the other wind lines.

H$\alpha$ shows consistent redshifted and blueshifted emission components at velocities of $\pm60$ km s$^{-1}$. The blueshifted emission appears in the earliest epochs despite the significant wind absorption, and toward later epochs rises to equal the strength of the redshifted component (see also the 2019 epochs of Park et al. 2020). The red- and blueshifted emission features are also apparent in the profiles of the Ca II IRT lines in later epochs. On closer inspection, the feature may be discernable as an inflection point in the profiles of the Ca II IRT lines in early epochs, as well as the H$\beta$ line profiles. Strangely, the peaks of the emission in the Ca II IRT lines are consistent with the locations of the peaks in H$\alpha$ in the data, but after removing the disk contribution, the peaks shift to lower velocities ($\pm40$ km s$^{-1}$).

The locations of the peaks of these emission components are approximately consistent with the Keplerian velocity at the innermost radii of the accretion disk. We interpret that the emission arises from a hot boundary layer, which may be shocking against the stellar photosphere.

The deep, narrow, low-velocity absorption component may trace an outflow with a wide opening angle, as has been observed in T Tauri systems (e.g., Whelan et al. 2021). These low-velocity absorption features are typically attributed to disk winds in T Tauri systems, which may indicate we are also seeing a slower disk wind.

In summary, we have demonstrated in this subsection that several of the strong lines traditionally used to diagnose inflowing and outflowing gas in YSOs can have contributions from disk absorption in FU Ori–type systems. Subtraction of the disk component reveals both inflowing possible boundary layer material, and outflowing slow and fast winds.

### 5.2. Broad Central Absorption

The evolution of the especially high-EP lines mentioned briefly in Section 4 (due to Si II, Si I, O I, Ca II, and C I) is shown in Figure 12. The lines follow a similar evolution to that of the HP lines, indicating the lines all trace a rapidly cooling hot component of the system. However, unlike the HP lines, these lines are significantly stronger in our spectra than in the disk model (especially in the earliest epochs) and can be clearly seen in the residuals in Figure 2. The profile shapes are also very similar in the residuals after we subtract the disk model. The significant differences in the depths and profile shapes for these lines from the predictions in our disk model indicate the majority of the absorption traced by them is not accounted for in our model. We note that the residual depths are much smaller in the later epochs, indicating the absorption excess relative to the disk model has almost full disappeared by the 2017 epoch.

The component traced by the high-EP lines appears to be somewhat dynamically different from the disk. The lines show a largely round-bottomed profile, almost consistent with profiles of high $v \sin i$ stellar features. The wings of the lines show relatively consistent 60 km s$^{-1}$ broadening, which may be attributed to the disk broadening seen in other features. However, the line core shows an initial 40–50 km s$^{-1}$ broadening which decreases as the line weakens, narrowing to only 20–30 km s$^{-1}$. The lines appear centered at the systemic rest velocity, implying they are tracing a particularly slow, albeit hot, outflow.

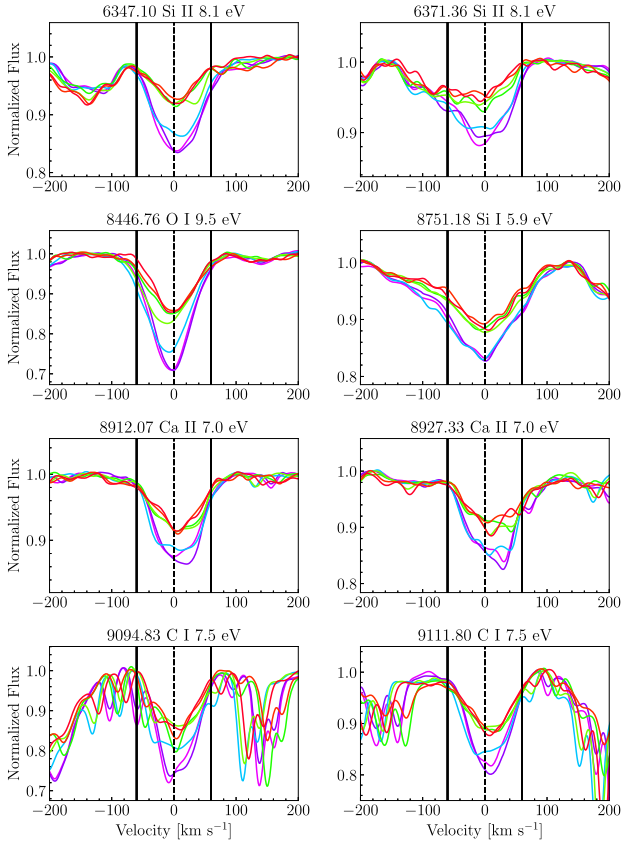

**Figure 12.** High-EP lines in the HIRES spectra of V960 Mon, smoothed with a $\sigma = 5$ pixel Gaussian for clarity. Epochs are shown in different colors, where redder indicates later epochs when the target is dimmer and, according to our disk model, cooler. The black vertical lines mark $\pm 60$ km s$^{-1}$, the estimated $v_{max}$ at outburst, as a reference for the line widths. The C I $\lambda$9095 line is significantly contaminated by telluric absorption, but its similar evolution to the other lines is still apparent.

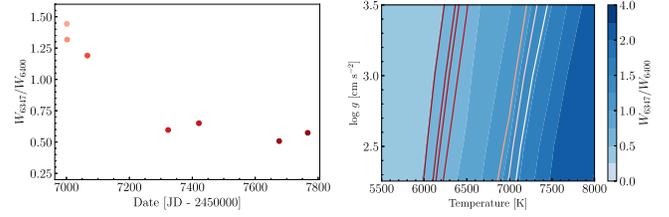

**Figure 13.** The EW ratio of the Si II $\lambda$6347 line to the Fe I $\lambda$6400 line for the HIRES spectra, showing that the ratio decreases in time and is consistent with a decreasing temperature. The color schemes in the left and right panels both show lighter red for larger EW ratios. Left: the time series of the EW ratios. Right: the EW ratios of the same lines as measured in the PHOENIX grid (blue contoured background) and in the HIRES spectra (red contours).

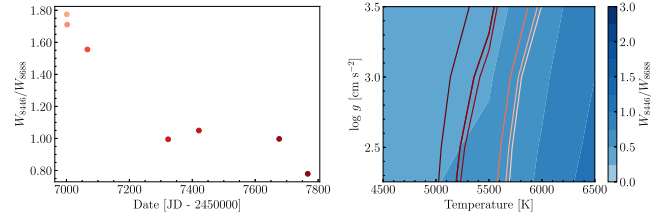

**Figure 14.** The EW ratio of the O I $\lambda$8446 line to the Fe I $\lambda$8688 line for the HIRES spectra, showing that the ratio decreases in time and is consistent with a decreasing temperature. The color schemes in the left and right panels both show lighter red for larger EW ratios. Left: the time series of the EW ratios. Right: the EW ratios of the same lines as measured in the PHOENIX grid (blue contoured background) and in the HIRES spectra (red contours).

To quantify the evolution of these high-EP lines better and to try to understand the implications for the temperature conditions in this additional component, we again use EW ratios like those described in Section 4. For our investigation, we choose two lines: Si II $\lambda$6347 and O I $\lambda$8446. These lines are high EP (8.1 eV and 9.5 eV, respectively), show a dramatic decrease in strength over time, and are well isolated from other photospheric lines and telluric lines. For the denominators in the line ratios, we choose the Fe I $\lambda$6400 and $\lambda$8688 lines, both of which show almost no time variability.

The time series of the $\lambda$6347/$\lambda$6400 and $\lambda$8446/$\lambda$8688 ratios are shown in Figures 13 and 14. Both sets of ratios show an initial rapid fade, similar to that of the HP lines, followed by a plateau, closely following the structure of the V-band lightcurve. The $T_{eff}$ versus log $g$ contour plot for the $\lambda$6347/$\lambda$6400 ratio shows that the line ratio is initially consistent with a $\sim$7000 K temperature atmosphere and as the target fades the ratio is more consistent with that seen in $\sim$5800 K atmospheres.

The contour plot for the $\lambda$8446/$\lambda$8688 ratios also shows a similar evolution, where the ratio is initially consistent with $\sim$5700 K atmospheres and evolves to be more consistent with $\sim$5200–5300 K atmospheres. Both sets of ratios indicate then that this excess component is cooling as the target fades. They also show, however, that different wavelengths are indeed tracing regions with different temperatures. This is consistent with the expected $T_{eff}(\lambda)$ relation expected for accretion disks and indicates that the continuum at 6400 Å arises from a warmer region of the disk than that at 8400 Å. We can use the temperatures described above and the wavelengths of the line ratios to find that at outburst, $dT/d\lambda \sim 0.65$ K Å$^{-1}$ whereas at the 2017 epoch, $dT/d\lambda \sim 0.45$ K Å$^{-1}$.

Taking the temperature from the $\lambda$6347/$\lambda$6400 ratio at outburst to be $\sim$7000 K and the temperature from the $\lambda$8446/$\lambda$8688 ratio to be $\sim$5700, we can also estimate that the lines arise from the $r \sim 2\,R_*$ and $r \sim 3\,R_*$ annuli respectively. That would in turn mean an average $\frac{dT}{d\lambda} \sim 0.62$ K Å$^{-1}$ across the red range of the optical spectrum.

We now turn our attention to some cooler lines, namely moderate-EP lines of Fe II that also show broad excess absorption relative to the disk model. Similarly to the high-EP lines, the Fe II lines (shown in Figure 15) also decrease in strength significantly as V960 Mon fades. The effect is not as dramatic as that seen in the high-EP lines such as Si II $\lambda$6347, but it is notable. The Fe II lines are also much stronger than predicted in the disk models, as can be seen for the Fe II $\lambda$5316 and $\lambda$5362 lines in Figure 2. These two facts imply the Fe II lines may trace the same excess component as the high-EP lines.

However, there are also Fe II lines, especially those blueward of 5200 Å, that show almost no time evolution. The Fe II lines which most closely follow the evolution of the high-EP lines in Figure 12 are those between 5200 and 7000 Å having EPs between 3.0 and 4 eV. We note that these Fe II lines are also typically gravity sensitive in single-atmosphere models, but when varying the log $g$ of our disk model, we do not see significant gravity sensitivity.

In summary, in this subsection we have demonstrated that there is a family of lines with deep broad profiles in V960 Mon that have distinctly deeper depths, and very different broadening shapes relative to the disk model. These lines come exclusively from high- and intermediate-EP species. We speculate that they originate in the hot components of the wind, near its point of origin in the disk.

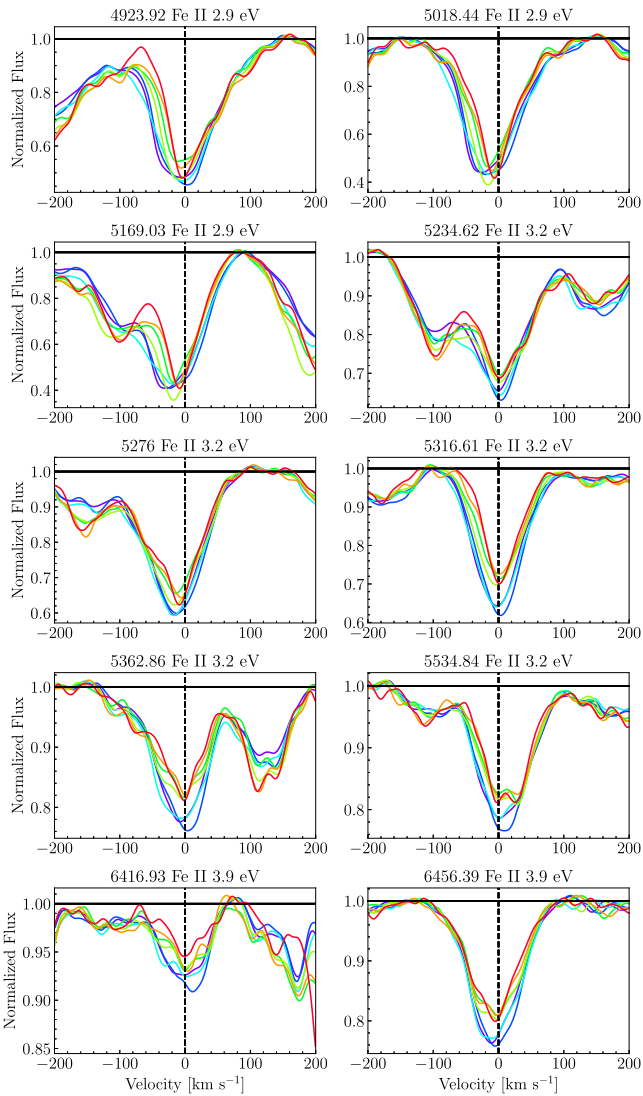

**Figure 15.** Fe II lines in the HIRES spectra. Notice their behavior is similar to the high-EP lines shown in Figure 12, rapidly shrinking over time. This effect is strongest in the lines with higher EP, indicating sensitivity to the hot wind component. The Fe II lines here seem to trace the disk behavior at lower EPs and bluer wavelengths, and trace the wind at higher EPs and redder wavelengths.

### 5.3. Narrow Central Absorption

Several lines in the spectra at later epochs show an excess narrow absorption feature centered at the systemic velocity. The feature deepens over time, but does not broaden, maintaining a relatively consistent ∼20 km s$^{-1}$ half-width measured from the intersection of the narrow component with the disk component. The lines exhibiting this feature most prominently are shown in Figure 16.

Narrow central absorption lines span most of the optical range covered by the HIRES spectra but seem especially prominent between 4900 Å < λ < 8400 Å. The lines with excess narrow absorption components predominantly occupy a narrow range of somewhat high EPs: 2.5–3.5 eV, such as Ca I

(shown in detail in Figure 4) and Fe I though there are a few at lower EPs, including the Ba II and Li I features (all shown in Figure 16). A comparison with the PHOENIX model spectra shows the depths of many of these features are consistent with $T_{\rm eff} \sim$ 5000–7000 K atmospheres. The lines are also generally temperature sensitive and grow deeper in the PHOENIX atmosphere models as $T_{\rm eff}$ is lowered from 7000 to 5000 K.

The somewhat high temperatures at which we expect these features, the fact that they appear at the systemic velocity, and their growth as the high-EP lines discussed in Section 5.2 shrink, imply a connection between the two families of lines. As mentioned in Section 5.2, the high-EP lines shown in Figure 12 are all initially much deeper than predicted by our disk model, but over time their strength decreases to be more consistent with the models at later epochs. We have also shown that the evolution of the lines is consistent with this hot excess component cooling over time.

The low velocity of the absorption may be consistent with absorption by a slow outflow at a distance of $r \sim$ 4–5 $R_{\rm inner}$. This region would be consistent with broadening <35 km s$^{-1}$ as we see in this excess. If the outflow cools as the disk cools, the highest-EP levels may depopulate and fill the 2.5–3.5 EP levels, contributing to increased absorption in those species.

When we isolate the high- and low-EP excess absorption by subtracting the disk model from the data, we can get a spectrum of this nondisk component. Parts of the residual spectrum are shown in Figure 2, but we reproduce it in better detail in Figure 17. In Figure 17, the residual spectrum is shown compared to two log $g = 1.5$ PHOENIX atmospheres, one with $T_{\rm eff}$ = 9000 K and another with $T_{\rm eff}$ = 7000 K. For better comparison with the line widths we see in the residuals, we broadened both models to $v \sin i$ = 30 km s$^{-1}$.

The outburst epoch residual spectrum possesses features that are quite similar to those seen in the 9000 K spectrum (e.g., Si II λ6347 and λ6371, Fe II λ6456, and O I λ8446). As the target fades, features that are similar to those in the 7000 K spectrum (e.g., Ca I λ6439, λ6449, and λ6463) grow stronger. The upper left panel of Figure 17 shows this well, where there are several lines in the 4750–4850 Å range that are not in the 9000 K atmosphere but can be seen in the residual spectrum. The lower left panel of Figure 17 shows the same fact in another wavelength range. In this panel, the Fe II λ6456 line appears in both the 9000 and 7000 K atmospheres, where the Ca I λ6439 and λ6449 lines are not expected in the hotter atmosphere. They are indeed generally weaker initially. However, as the target fades, the Ca I features in the residual spectrum deepen until they are consistent in depth with those seen in the 7000 K spectrum. The inverse case is shown in the upper right panel of the figure, where the Si II features weaken over time, which is again what we may expect from a residual component cooling.

In summary, in this subsection we have demonstrated that there is a family of narrow absorption lines in V960 Mon that grows in strength over time against the "continuum" of the disk photosphere, which produces disk-broadened profiles in many of the same lines. We speculate that the lines may trace a slow-moving, cooler component of the wind that was initially traced by the high-EP excess absorption discussed in Section 5.2.

### 6. Forbidden Emission

The [S II] λ6731 emission feature was reported in the spectrum of V960 Mon by Park et al. (2020), and they discuss the growth of the feature at later epochs. We recover this feature and its growth in our HIRES spectra. We also find the [O I] λ6300 and [N II] λ6583 emission features. The maximum normalized flux of the [O I] λ6300 line is a bit greater than that of the [S II] λ6731 line. The [O I] and [N II] fluxes also increase relative to the continuum at later epochs. The features as seen in the HIRES spectra are shown in Figure 18.

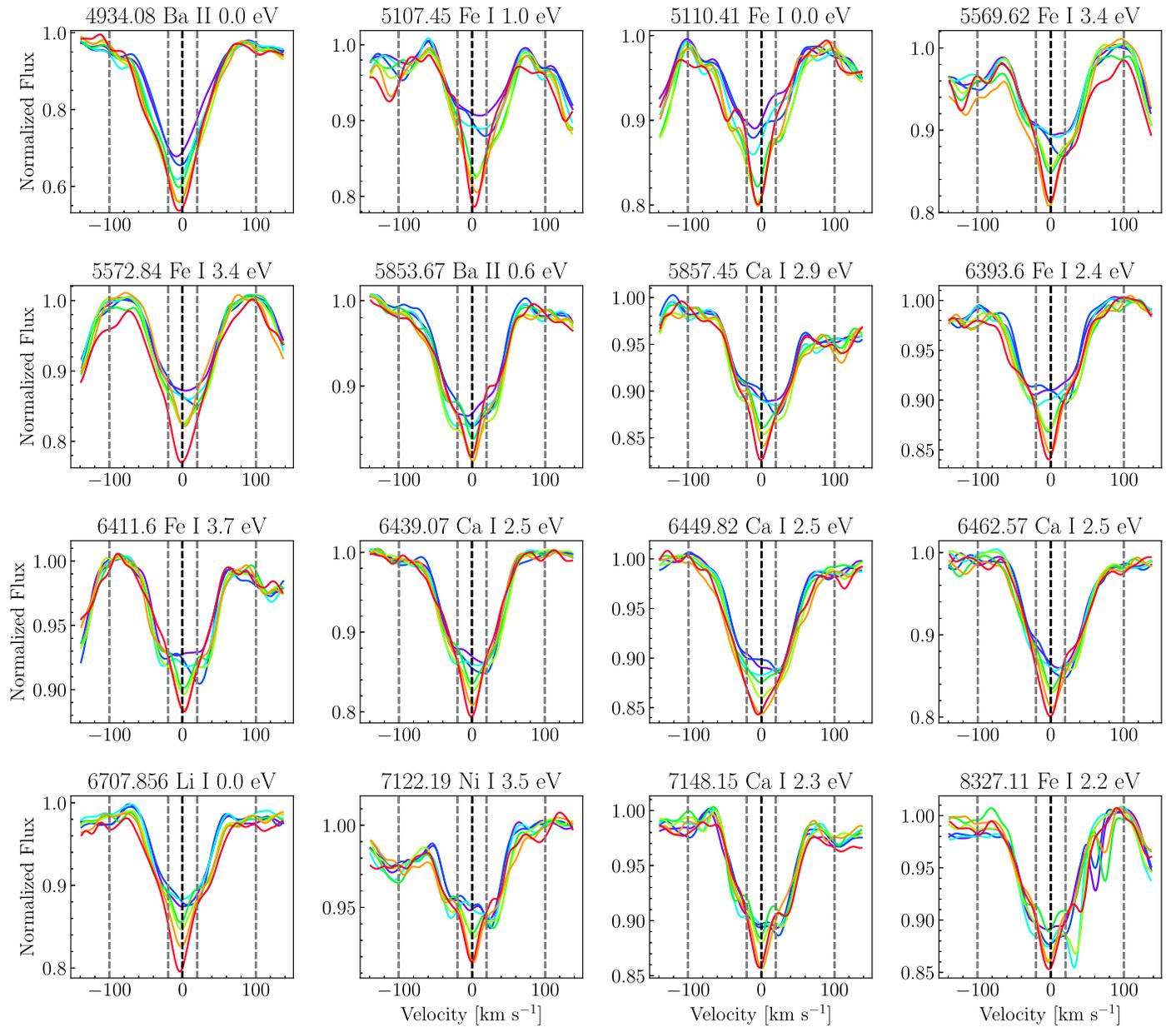

**Figure 16.** Lines in the HIRES spectra which show the central absorption feature shown as a time series progressing from blue to red color. Notice that across the range of lines and EPs, the narrow feature shows a consistent width ∼20 km s$^{-1}$ and is centered at 0 km s$^{-1}$. The feature grows as the target fades, absorbing against the disk continuum to additional depth that is almost equivalent to the depth of the disk contribution of many of the lines

The weaker components of these two doublets, the [S II] line at 6716 Å and [O I] line at 6363 Å, are tentatively identifiable in the latest epoch but are barely 1% above the continuum. The doublet of the [N II] line falls in between orders in the HIRES spectra. Detecting the 6716 and 6363 Å emission in the data is complicated by line blending with nearby photospheric lines, namely the Fe I λ6715 and λ6362 absorption lines, which are sufficiently broadened to blend with the neighboring forbidden emission features. Fortunately, the Fe I lines can be removed using our high-dispersion models of the HIRES spectra. Looking at the residuals, the features appear clearly and show similar structure to the 6731 and 6363 Å features. The stronger 6731 and 6300 Å lines are also clearly seen in the residuals and retain the amplitude we see in the data.

The line profiles show a velocity structure different from that we see in the absorption features. The emission is predominantly blueshifted in all three main lines, indicating they are tracing outflows from the FU Ori object. They also show multiple velocity components, the velocities of which differ slightly from line to line. To measure the widths and central velocities of the components, we use the `optimize` toolkit in `scipy` to do least-squares fitting of a sum of two Gaussian functions to the line profiles. The fits are quite noisy because the detections of the lines are weak and not all epochs are well described by a sum of two Gaussians. However, the best-fit parameters provide a general picture of the composition of the forbidden emission.

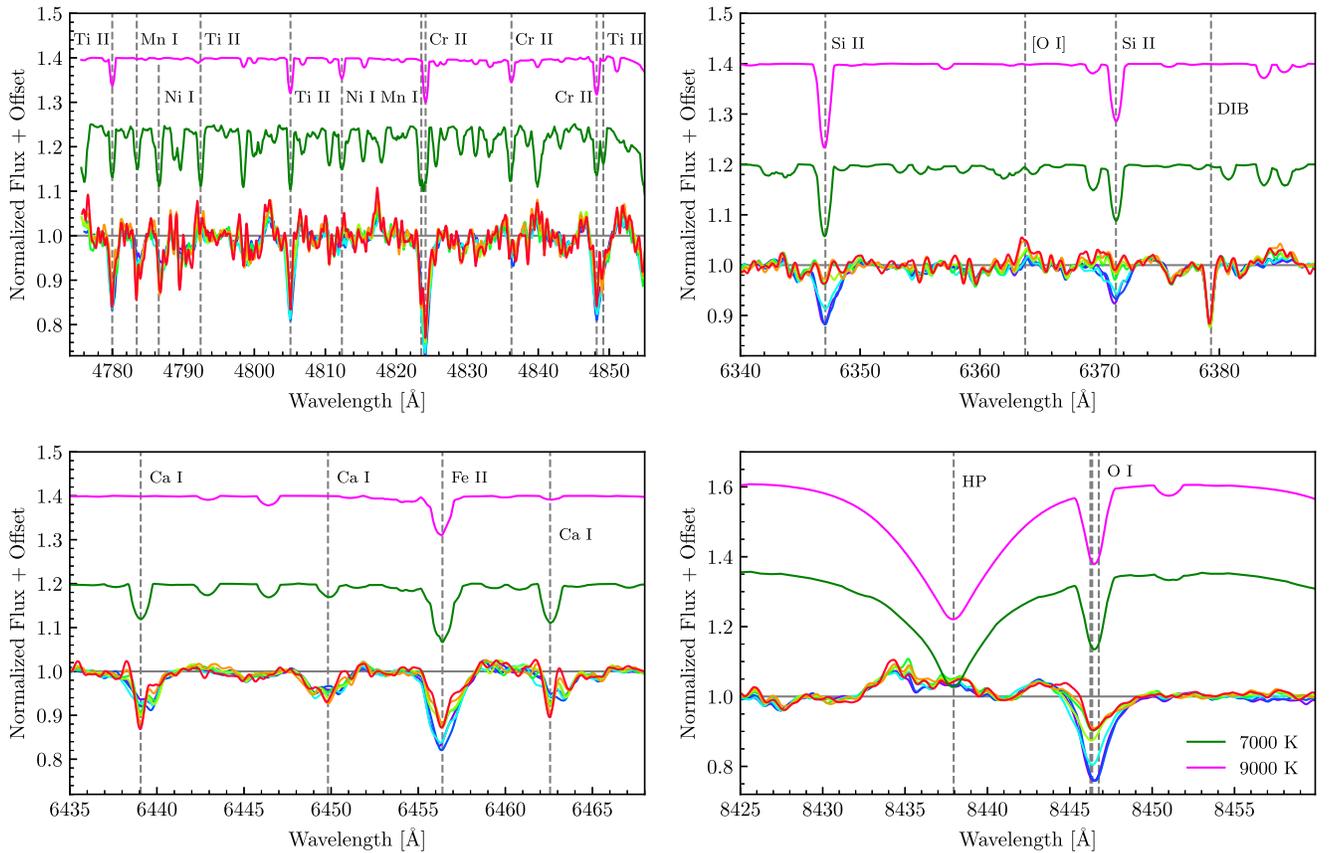

**Figure 17.** Selected regions of the residuals of our fits to the HIRES spectra, compared with log $g$ = 1.5, $T_{eff}$ = 7000 K (green) and $T_{eff}$ = 9000 K (magenta) PHOENIX models. Notice the residual features are consistent with the features seen in these hot model atmospheres, with a tendency for later epochs to more closely resemble the 7000 K atmosphere. This may indicate that the component of the system probed by the residuals is cooling as the disk cools. The H P lines in the lower right panel are strongly affected by Stark broadening in the PHOENIX atmospheres. That we do not see this in the disk model or the data indicates that the contribution from the narrower and weaker H P lines in cooler atmospheres is necessary to match the data.

The [S II] λ6731 line has two distinct peaks, one at essentially systemic rest velocity and one at −30 km s$^{-1}$. Both peaks have an HWHM of ∼15 km s$^{-1}$. The [O I] line also has a peak at systemic rest velocity but its blueshifted component is closer to −50 km s$^{-1}$. Its components are also both broader, with HWHM values of ∼20 km s$^{-1}$. The [N II] line seems dominated by a single component at −30 km s$^{-1}$, which has an HWHM of ∼25 km s$^{-1}$, though there may also be a weak rest-velocity component.

The $v \sin i = -30$ km s$^{-1}$ peak is consistent with the velocity of the narrow absorption component in the wind lines described in Section 5.1.

## 7. Discussion

The accretion disk model presented in Paper I successfully reproduces most of the variability seen in the spectra, as can be seen in the residuals presented in Figure 2. There are also many features in the residuals that are not captured by the disk model, either because they are classic wind tracers, such as Hα, the Ca II triplet, and the Na D lines, or because they trace some other, unmodeled component of the system (see Section 5).

### 7.1. Disk Absorption and the Gravity and Temperature Dependencies

The observed high-dispersion spectra are reasonably well reproduced by models of an accretion disk photosphere. However, great care is needed in understanding and disentangling the effects of temperature and surface gravity in the disk, from the wind components.

The lines that are typically used as gravity indicators cannot be used to determine the gravity of FU Ori objects because they are often sensitive to outflows and are therefore dominated by outflow absorption. This is the case for the Na D lines and many other highly gravity-sensitive features (see Section 5.1 for a discussion of these and other wind lines). We therefore must rely on other gravity indicators in the spectrum of V960 Mon, in the weaker optical atomic lines that are known to trace the disk.

To explore the gravity sensitivity of the absorption lines in the spectrum, we computed a grid of high-dispersion models using the outburst epoch best-fit parameters and in each we fixed the log $g(r)$ to be one of 0.5, 1.0, 1.5, 2.0, 2.5, 3.0, or 3.5. We then compared the change in the spectra due to the variation of log $g$ with the time evolution of the HIRES spectra.

Although we do see some evolution in the high-EP lines discussed in Section 5.2, they remain at least a factor of two deeper in the data than in the models, indicating once again that they are likely tracing excess absorption not accounted for in our disk model. The Fe II lines are also much deeper in the data than in any of the models, similarly indicating that they may also in part trace this hot excess absorption component. Hartmann & Calvet (1995) show that the Fe II lines in other FU Ori objects, such as FU Ori itself or V1515 Cyg, are consistent with wind absorption models. We do not see blueshifted

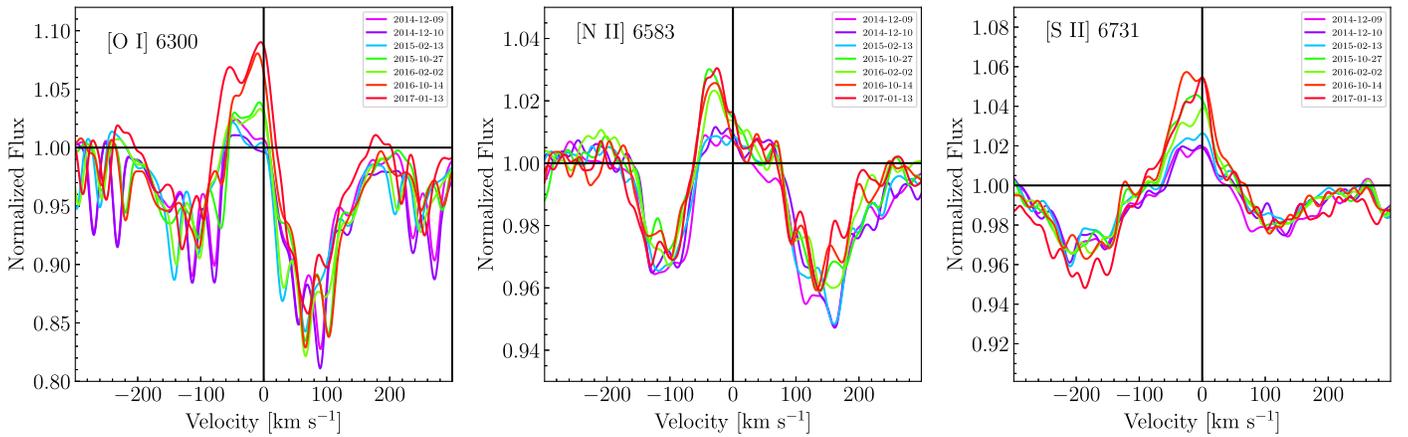

**Figure 18.** The [O I] λ6300, [N II] λ6583, and [S II] λ6731 emission lines in the HIRES spectra. All three features are blueshifted, with emission components ranging from −60 km s$^{-1}$ to −10 km s$^{-1}$. The emission in the [O I] λ6300 and [N II] λ6583 lines is mostly at higher velocities, whereas the [S II] λ6731 line shows both a −40 km s$^{-1}$ component and a nearly 0 km s$^{-1}$ component.

absorption in these profiles, but they may trace a slower outflowing component.

For the other gravity-sensitive lines, such as Ti I λ8435, explaining the line evolution seen in the HIRES spectra with only a change in the disk gravity would require a very large log g increase (at least ∼2 dex). In general, we find that the overall spectrum is more consistent with the lower gravity models (log g ∼ 1.5) than models that use a g(r) profile. However, this should also be tested with the gravity-sensitive features in the near-infrared (NIR), such as the CO (2–0) and (3–1) band heads.

Ultimately, we find that it is critical to look at the disk-integrated model spectrum to study the gravity sensitivity of different features, rather than looking at them atmosphere by atmosphere. The gravity sensitivity of a given feature is a function of the effective temperature of the atmosphere and may not extend across broad temperature ranges. Therefore, the temperature blending in the full integrated disk spectrum may counteract or weaken the gravity sensitivity of lines like Ti I λ8435.

The evolution of the Fe I and Ca I lines, on the other hand, is insensitive to log g but consistent with a ∼1000–2000 K decrease in $T_{\rm eff}$. The lower-EP Fe I lines are especially temperature sensitive and grow rapidly as the target fades. The higher-EP Fe I lines do not vary as significantly initially, but consistently show a growing narrow central absorption feature at later times. We also see a similar pattern in the Ca I features. The Ca I λ6439 and λ6462 features show little evolution except for growth of the narrow central absorption.

The relative lack of the gravity sensitivity and significant temperature sensitivity of these two species in the temperature ranges we expect (5000–7000 K) are indicative that the narrow central absorption is not tracing a higher-gravity component of the system but rather one that is cooling.

### 7.2. Wind Features

The wind lines shown in Section 5.1 show evidence of a multicomponent outflow that evolves over the course of the postoutburst fading. The primary components we identify are a fast-moving (−200 km s$^{-1}$) component traced by the Hα and Hβ lines, a slower component (−50 to −30 km s$^{-1}$) traced initially by Hα, Hβ, and the Ca II IRT, and a very slow component (−10 to 0 km s$^{-1}$), traced by all lines (the Na I D doublet in all epochs and the other lines at later epochs). The velocities probed by the emission components in the [O I], [N II], and [S II] forbidden emission lines discussed in Section 6 are also consistent with some of the slower components seen in the wind lines.

The fastest component of the wind that is visible in the −200 km s$^{-1}$ absorption of the Hα and Hβ lines is similar in shape to that seen in the Na I D and Hβ lines of other FU Ori objects like V1057 Cyg and FU Ori (Hartmann & Calvet 1995). This absorption is consistent with a high mass-outflow rate, $\dot{M}_{\rm out}$, during the outburst, as demonstrated in models of disk winds during FU Ori outbursts (Calvet et al. 1993; Milliner et al. 2019).

As the target fades, however, we see this high-velocity component disappear, indicating the $\dot{M}_{\rm out}$ is much lower. Following the estimated $\dot{M}_{\rm out} \sim 0.1\, \dot{M}_{\rm acc}$ described in Calvet et al. (1993), we would expect that the high-velocity component would remain relatively deep, because we only predict a 40% decrease in $\dot{M}_{\rm acc}$. To preserve the $L_w/L_{\rm acc}$ they describe (which is also supported by Zhu et al. 2020), we would require a much more massive wind corresponding to the apparent velocity decrease, which would likely make it optically thick.

The slow component of the outflow, visible in Hα, Hβ, the Na D lines and the Ca II IRT, ranges from −10 to −30 km s$^{-1}$. In the Na I D lines in particular, this component is saturated, indicating it may be a very massive outflow and absorbs against a large area of the disk from which the visible continuum arises. In the other lines, the component is initially more blueshifted (up to −50 km s$^{-1}$) but slows to −10 km s$^{-1}$ over time. The persistence of the component in Na I, and its appearance in the Hα, Hβ, and Ca II IRT lines at later epochs suggests the outflow is present during the course of the initial outburst, throughout the fade and into the plateau. It may be that the faster outflow traced by the other lines obscures the slower component we see in Na I.

This absorption is also consistent in velocity with the −30 km s$^{-1}$ component we see in the [S II], [O I], and [N II] forbidden emission (Section 6). This may be evidence of a disk wind similar to that seen in T Tauri stars (Whelan et al. 2021), though it may be much more massive.

The forbidden emission lines shown in Section 6 may allow us to differentiate between different outflow components. The [O I] λ6300 emission line has the highest critical density of the three emission features ($n_c \sim 2 \times 10^6$ cm$^{-3}$) and also shows the highest-velocity emission feature ($-50$ km s$^{-1}$). This may be emission from a region closer to $R_{\rm inner}$, where the wind density and escape velocity may be higher. The [S II] emission, which has the lowest critical density ($n_c \sim 2 \times 10^4$ cm$^{-3}$), may in turn probe a region further from $R_{\rm inner}$, and therefore has a lower velocity.

Both the [S II] and [O I] features also have $v \sim 0$ km s$^{-1}$ emission features at similar strengths to their blueshifted emission. This very-low-velocity emission is consistent with the low-velocity components of the forbidden emission observed in T Tauri disks, which are believed to be trace slow-moving disk winds (as opposed to the "high-velocity components" that are usually attributed to jets; see Pascucci et al. 2023 for a review). It is possible that due to the increase in $T_{\rm eff}$ and $L_{\rm bol}$ of the source irradiating the outer disk, the wind that is launched may be more massive than those in T Tauri systems. This would be consistent with the finding by Cruz-Sáenz de Miera et al. (2023) that FU Ori outflows observed in cooler molecular emission are indeed be more massive than those of T Tauri systems.

Using the dereddened fluxes of our SED models at each HIRES epoch, we are able to estimate the total flux in the emission lines. Though the target fades significantly over time, we find that the fluxes of the [O I] λ6300 and [S II] λ6731 lines are relatively constant, perhaps increasing slightly over time. The measured median fluxes for the lines are $\log \lambda F_\lambda(6300) = -14.2 \pm 0.34$ and $\log \lambda F_\lambda(6731) = -13.9 \pm 0.11$ erg s$^{-1}$ cm$^{-2}$. Using the distance to the target of 1120 pc, we estimate line luminosities of $\log(L_{6300}/L_\odot) = -3.66 \pm 0.34$ and $\log(L_{6731}/L_\odot) = -3.36 \pm 0.11$. For the [O I] λ6300 line, these values are almost 1.5 dex greater than the $\log(L_{6300}/L_\odot) \sim -5$ reported for several classical T Tauri Stars by Fang et al. (2018).

The Hα and Ca II IRT lines show a strong redshifted emission component, centered at $+60$ km s$^{-1}$, which grows stronger over time. Toward later epochs, the strong absorption features in Hα and Ca II IRT also become dominated by blueshifted emission at $-60$ km s$^{-1}$. The Hα profiles shown in Park et al. (2020) for their 2018 epochs continue to show the double-peaked Hα emission, and the peaks remain at $v \sim \pm 60$ km s$^{-1}$. The velocities of the emission peaks are consistent with the $v_{\rm kep} \sin i$ of $R_{\rm inner}$. This may be evidence of emission from the innermost radius of the disk, tracing the accretion boundary with the star.

For the broad/shrinking and narrow/growing low-velocity absorption excess components, we believe these can be attributed to some sort of excess, outflowing material above the disk atmosphere. The rest-velocity absorption profile is consistent with observations of disk winds through [O I] emission profiles in face-on circumstellar disks (Fang et al. 2023). The notion of a hot excess component cooling and a cooler component appearing is also consistent with the behavior in the coronal X-ray emission of the system (Kuhn & Hillenbrand 2019).

## 8. Conclusion

In Paper I, we presented a means of modeling the SED of the V960 Mon system at outburst using information from our HIRES spectrum at outburst to constrain the SED fit and break some existing degeneracies between the physical parameters in the model. We also presented a means of estimating the $\dot{M}$ and $R_{\rm inner}$ of the system for subsequent epochs as the system faded postoutburst.

In this work, we used the disk parameters in Paper I to construct high-spectral-resolution models at each observational epoch in our high-spectral-resolution time series data set, in order to understand the evolution of both disk and nondisk components of the V960 Mon system better. We have shown:

1. Our high-resolution model disk spectrum accurately reproduces the evolution of disk absorption features across the 4000–9000 Å range of the HIRES spectra during the postoutburst fading of the V960 Mon system.
2. The HIRES spectra show evidence of temperature evolution that is consistent with our predicted SED evolution of the system.
3. We are able to isolate absorption and emission from nondisk components of the V960 Mon system, including a strong multicomponent outflow, by subtracting our model disk spectrum from the HIRES spectra.
   (a) We detect [O I], [N II], and [S II] emission that is consistent with the multicomponent forbidden emission from classical T Tauri disk systems, though these likely trace much more massive outflows because they tend to be as bright shock emission from jets.
   (b) At outburst, the spectra show a very massive, high-velocity outflowing wind that weakens and largely disappears as the target fades.
   (c) Several high-EP lines in the spectra show strong and broad rest-velocity excess absorption, which weakens as the target fades. Correspondingly, the lower-EP lines show a narrow rest-velocity excess that strengthens as the target fades. We interpret this as evidence of a slow-moving, cooling outflow in the system.

Further high-resolution follow up of the system will be crucial for understanding how it compares at later epochs to the older, more mature FU Ori outbursts that have been well studied.

## Acknowledgments

We thank the Association of Amateur Variable Star Observers for their dedicated high-cadence sampling of the postoutburst lightcurve of this target, which we have reproduced in Figure 1, and discussed for multiple bands in Paper I.

We thank the anonymous referee for the detailed comments which helped improve the manuscript.

## Appendix A
## Optical HWHD Measurements and the Keplerian Disk Model

In past studies of FU Ori objects, HWHD or FWHD measurements of spectral lines have been used to estimate the location in the disk from which the spectral lines arise (Herbig et al. 2003; Park et al. 2020). However, repeated investigations of line widths in FU Ori objects have found little variation in them as a function of wavelength, particularly in the optical.

Historically, this observation has been raised as evidence against the canonical $T \propto R^{-3/4}$ and $v \sim v_K$ models used. This is because the line width measurements are assumed to follow a velocity profile similar to that shown in Figure A1. The velocity profile is derived by taking the $v_{\rm kep}$ at the luminosity-weighted mean radius of each wavelength bin in our outburst SED model.

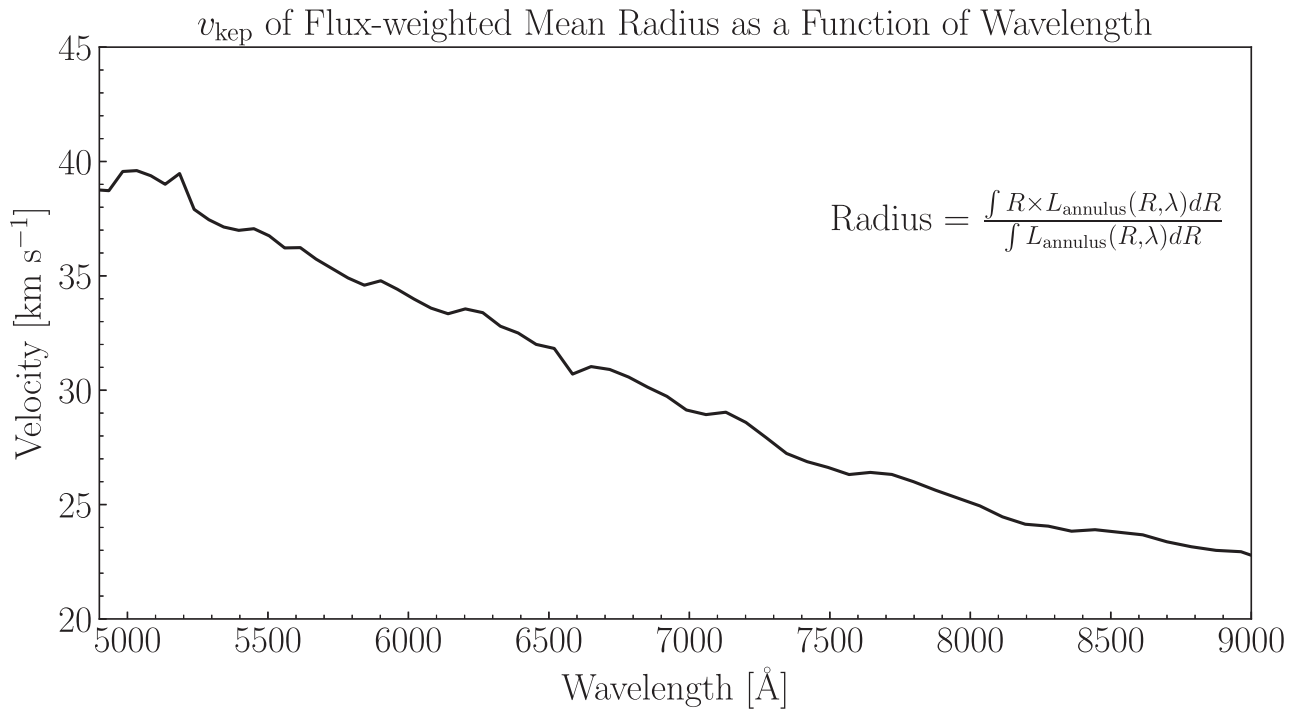

**Figure A1.** The Keplerian velocity at the luminosity-weighted mean radius of each wavelength bin in the outburst spectrum model. Spectral absorption features are visible in the velocity profile because annuli with greater total flux at the slightly greater radii are needed to compensate for the absorption.

Contrary to these expectations, however, we find that measurements of the HWHD in our Keplerian, thin disk, $T \propto R^{-3/4}$ disk model are relatively consistent with those we measure in the HIRES spectra (see Section 3.3). Though there is some slope in the measurements, tending toward narrower lines at redder wavelengths, the correlation is not significant. Furthermore, the widths of individual spectral lines is not sufficient to estimate the radial locations of the annuli from which the lines arise due to the differential broadening effect described in Section 3.2. The limited utility of HWHD measurements of spectral lines in the visible range is also pointed out in Zhu et al. (2009), where the authors argue that much broader wavelength ranges must be used. Zhu et al. (2009) find that lines around 5 $\mu$m do indeed have significantly narrower profiles than those in the visible.

## Appendix B
## Attempts to Fit the TiO Band Heads

We generally find some inconsistency between the predictions of the disk model in the NIR and the data, particularly in its ability to match molecular features. In Paper I, we show that our model is capable of reproducing the spectrophotometry of V960 Mon in the NIR near the outburst peak. However, the model fails at later epochs when the H$_2$O $H$-band features are significantly deeper. The difficulty in matching molecular absorption consistently across our disk model begins near the 8600 Å region, where the model predicts multiple strong TiO bands that should be visible between $I$ and $Y$ band, but they do not appear in the data.

The problem with the red/NIR TiO band heads in FU Ori objects is also mentioned by Herbig et al. (2003) in their attempts to use a disk model to model the spectra of FU Ori and V1057 Cyg. Their model uses only three stellar templates, chosen to represent three different temperatures in the disk, but they find the same problem, indicating that the issue does not arise from our choice of spectral model grid. Herbig et al. (2003) find that in both objects, the model predicts a very strong TiO $\lambda$8860 band head, though in FU Ori and V1057 Cyg (as in V960 Mon), only the HP line at $\lambda$8860 appears in the object spectra.

In our attempts to fit the TiO band head at 8860 Å we find that the band head is best replicated by having a disk that truncates at $R_{\rm outer} = 12\ R_\odot$, which is very compact, extending only to 6 $R_{\rm inner}$. A comparison between the band head in the data and the models with varying $R_{\rm outer}$ values is shown in Figure B1. An active disk component like that is too small because if it were truncated at such short radii, would have an outermost temperature of 4500 K. With this as the outermost (minimum) temperature of the active disk, no region of the active component would be cool enough to produce any of the NIR molecular features.

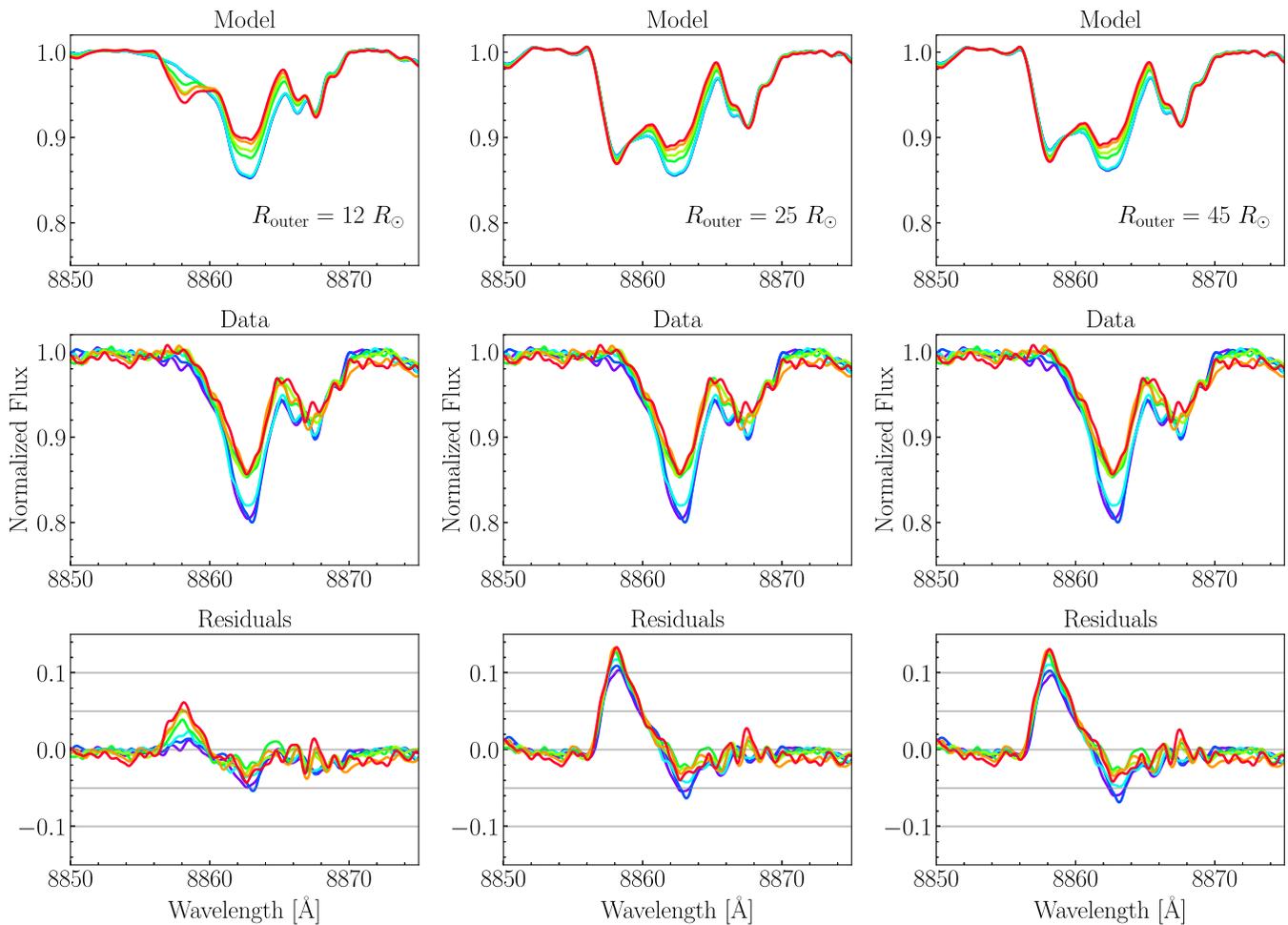

**Figure B1.** The TiO band head at 8860 Å in our models, the HIRES spectra, and the model residuals, shown for all of the HIRES epochs and for three different models. Left: models computed assuming $R_{\rm outer} = 12\,R_\odot$. Notice the time (and therefore $\dot{M}$) dependence of the TiO band head is much stronger for this model. Center: models using $R_{\rm outer} = 25\,R_\odot$. Already, there is little change in the TiO band head with time, and it clearly overpredicts the absorptions. Right: models using $R_{\rm outer} = 45\,R_\odot$. The TiO band head is essentially insensitive to $\dot{M}$ in these models and continues to overestimate the absorption we see.

## Appendix C
## All Orders of the HIRES Spectra and Residuals

Figure C1 shows all of the HIRES spectral orders from our seven observations, including those shown in Figure 2.

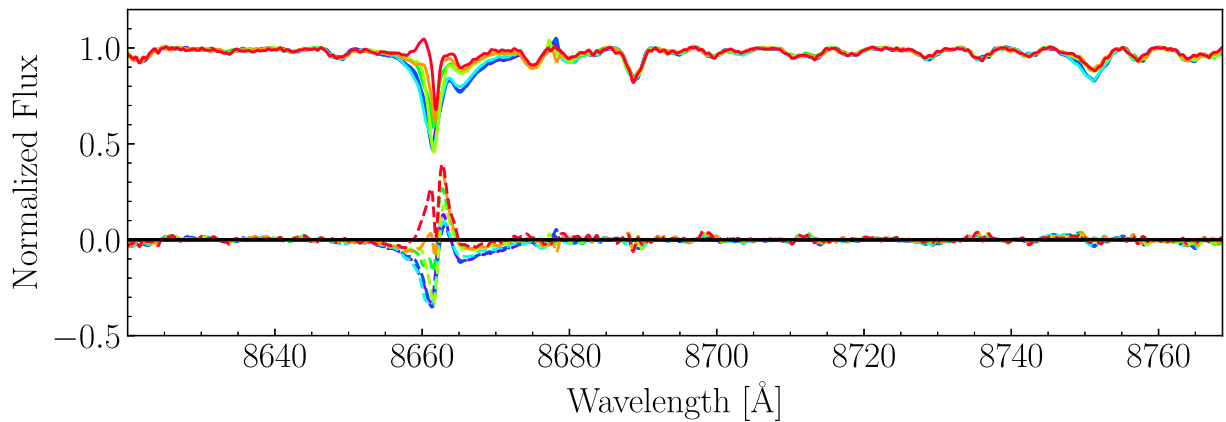

**Figure C1.** The V960 Mon visible range spectra for all seven HIRES epochs, shown with model residuals as dashed lines.

(The complete figure set (34 images) is available.)

### ORCID iDs

Adolfo Carvalho  https://orcid.org/0000-0002-9540-853X